\documentclass[twocolumn,twoside]{IEEEtran}
\usepackage{mathpazo}
\usepackage{times}

\usepackage{amsmath}
\usepackage{amsfonts}
\usepackage{latexsym}
\usepackage{amssymb}

\usepackage{upref}
\usepackage{theorem}
\usepackage{graphicx}
\usepackage{psfrag}
\usepackage{cite}

\usepackage[vlined,boxruled]{algorithm2e}
\SetKwInOut{Input}{input}
\SetKwInOut{Output}{output}
\SetKw{KwDownto}{down to}
\IncMargin{1.2em}
\LinesNumbered

\theoremstyle{plain}
\theorembodyfont{\normalfont\slshape}

\newtheorem{thm}{Theorem$\!$}
\newenvironment{theorem}{\begin{thm}\hspace*{-1ex}{\bf.}}{\end{thm}}

\newtheorem{clm}[thm]{Claim$\!$}

\newtheorem{lem}[thm]{Lemma$\!$}
\newenvironment{lemma}{\begin{lem}\hspace*{-1ex}{\bf.}}{\end{lem}}

\newtheorem{prop}[thm]{Proposition$\!$}

\newtheorem{cor}[thm]{Corollary$\!$}

\newtheorem{defn}[thm]{Definition$\!$}
\newenvironment{definition}{\begin{defn}\hspace*{-1ex}{\bf.}}{\end{defn}}

\newtheorem{xmpl}[thm]{Example$\!$}

\newtheorem{cnstr}{Construction$\!$}

\newtheorem{cnjc}{Conjecture$\!$}

\newcounter{enumrom}
\renewcommand{\theenumrom}{(\roman{enumrom})}

\makeatletter
\renewcommand{\@endtheorem}{\endtrivlist}
\makeatother

\makeatletter
\renewcommand{\thefigure}{{\@arabic\c@figure}}
\renewcommand{\fnum@figure}{{\bf Figure\,\thefigure}}
\makeatother

\newcommand{\cK}{\mathcal{K}}

\newcommand{\cM}{\mathcal{M}}

\newcommand{\mathset}[1]{\left\{#1\right\}}
\newcommand{\mathsetp}[2]{\mathset{#1 ~|~ #2}}
\newcommand{\abs}[1]{\left|#1\right|}
\newcommand{\ceilenv}[1]{\left\lceil #1 \right\rceil}
\newcommand{\floorenv}[1]{\left\lfloor #1 \right\rfloor}
\newcommand{\parenv}[1]{\left( #1 \right)}
\newcommand{\sparenv}[1]{\left[ #1 \right]}
\newcommand{\bracenv}[1]{\left\{ #1 \right\}}

\newcommand{\be}[1]{\begin{equation}\label{#1}}
\newcommand{\ee}{\end{equation}}

\renewcommand{\leq}{\leqslant}

\renewcommand{\geq}{\geqslant}

\renewcommand{\Bbb}{\mathbb}

\newcommand{\Cref}[1]{Co\-ro\-lla\-ry\,\ref{#1}}

\renewcommand{\Bbb}{\mathbb}

\newcommand{\N}{{\Bbb N}}
\newcommand{\R}{{\Bbb R}}
\newcommand{\Z}{{\Bbb Z}}

\newcommand{\ksucc}{\mathtt{Successor}_{\cK}}
\newcommand{\krank}{\mathtt{Rank}_{\cK}}
\newcommand{\kunrank}{\mathtt{Unrank}_{\cK}}
\newcommand{\lsucc}{\mathtt{Successor}_{\infty}}
\newcommand{\lrank}{\mathtt{Rank}_{\infty}}
\newcommand{\lunrank}{\mathtt{Unrank}_{\infty}}
\newcommand{\ind}{\mathtt{Ind}}
\newcommand{\sw}{\mathtt{sw}}
\newcommand{\suc}{\mathtt{Succ}}
\newcommand{\rn}{\mathtt{Rn}}
\newcommand{\unr}{\mathtt{UnR}}
\newcommand{\quo}[1]{``#1''}
\newcommand{\pttt}{\quo{push-to-the-top} }
\newcommand{\mvar}[1]{\text{\textit{#1}}}
\newcommand{\dw}[3]{\vphantom{#3}^{#1}_{#2}{\downarrow}#3}
\newcommand{\up}[3]{\vphantom{#3}^{#1}_{#2}{\uparrow}#3}

\outer\def\proclaim #1. #2\par{\medbreak
 \noindent{\bf#1.\enspace}{\sl#2\par}%
 \ifdim\lastskip<\medskipamount \removelastskip\penalty55\medskip\fi}

\begin{document}

\title{\textbf{Snake-in-the-Box Codes for Rank Modulation}}

\author{\large
Yonatan~Yehezkeally and
Moshe~Schwartz,~\IEEEmembership{Senior Member,~IEEE}
\thanks{Yonatan Yehezkeally is with the Department
   of Electrical and Computer Engineering, Ben-Gurion University of the Negev,
   Beer Sheva 84105, Israel
   (e-mail: yonatany@bgu.ac.il).}
\thanks{Moshe Schwartz is with the Department
   of Electrical and Computer Engineering, Ben-Gurion University of the Negev,
   Beer Sheva 84105, Israel
   (e-mail: schwartz@ee.bgu.ac.il).}
\thanks{
  This work was supported in part by ISF grant 134/10.}
}

\maketitle


\begin{abstract}
Motivated by the rank-modulation scheme with applications to flash
memory, we consider Gray codes capable of detecting a single error,
also known as snake-in-the-box codes. We study two error metrics:
Kendall's $\tau$-metric, which applies to charge-constrained errors,
and the $\ell_\infty$-metric, which is useful in the case of limited
magnitude errors. In both cases we construct snake-in-the-box codes
with rate asymptotically tending to $1$. We also provide efficient
successor-calculation functions, as well as ranking and unranking
functions. Finally, we also study bounds on the parameters of such
codes.
\end{abstract}



\section{Introduction}
\IEEEPARstart{F}{lash} memory is non-volatile storage medium which is
electrically programmable and erasable.  Its current wide use is
motivated by its high storage density and relative low cost.  Among
the chief disadvantages of flash memories is their inherent asymmetry
between cell programming (injecting cells with charge) and cell
erasure (removing charge from cells). While single cells can be
programmed with relative ease, in the current architecture, the
process of erasure can only preformed by completely depleting large
blocks of cells of their charge. Moreover, the removal of charge from
cells physically damages cells over time.

This issue is exacerbated as a result of the ever-present demand for
denser memory: smaller cells are more delicate, and get damaged faster
during erasure. They also contain less charge and are therefore more
prone to error. In addition, flash memories, at present, use
multilevel cells, where charge-levels are quantized to simulate a
finite alphabet -- the more levels, the less safety margins are left,
and data integrity is compromised. Thus, over-programming (increasing
a cell's charge-level above the designated mark) is a real problem,
requiring a costly and damaging erasure cycle.  Hence, in a
programming cycle, charge-levels are usually made to gradually
approach the desirable amount, making for lengthier programming cycles
as well (see \cite{BreGil08}).

In an effort to counter these effects, a different modulation scheme
has been suggested for flash memories recently -- rank
modulation \cite{JiaMatSchBru09}. This scheme calls for the
representation of the data stored in a group of cells in the
permutation suggested by their relative charge-levels. That is, if
$c_1,c_2,\ldots,c_n\in\R$ represent the charge-levels of $n\in\N$
cells, then that group of cells is said to encode that permutation
$\sigma\in S_n$ such that:
\[c_{\sigma(1)}>c_{\sigma(2)}>\ldots>c_{\sigma(n)}.\]
This scheme eliminates the need for discretization of
charge-levels. Furthermore, restricting ourselves to programming the
group of cells only by increasing the charge-level of a given cell
above that of any other cell in the group, over-programming is no
longer an issue. This operation was named in \cite{JiaMatSchBru09} as
a \pttt operation.

In addition, storing data using this scheme also improves the memory's
robustness against other noise types. Retention, the process of slow
charge leakage from cells, tends to affect all cells in a similar
direction \cite{BreGil08}. Since rank modulation stores information in
the differences between charge-levels rather than their absolute
values, data stored using it is more resilient to this sort of noise.

Gray codes using \pttt operations and spanning the entire space of
permutations were also studied in \cite{JiaMatSchBru09}. The Gray code
\cite{Gra53} was first introduced as a sequence of distinct binary
vectors of fixed length, where every adjacent pair differs in a single
coordinate.  It has since been generalized to sequences of distinct
states $s_1,s_2,\ldots,s_k\in S$ such that for every $i<k$ there
exists a function in a predetermined set of transitions $t\in T$ such
that $s_{i+1}=t(s_i)$ (see \cite{Sav97} for an excellent survey).
When the states one considers are permutations on $n\in\N$ elements
and the allowed transitions are \pttt operations,
\cite{JiaMatSchBru09} referred to such Gray codes as \emph{$n$-length
  Rank-Modulation Gray Codes} ($n$-RMGC's), and it presented such
codes traversing the entire set of permutations.  In this fashion, a
set of $n$ rank-modulation cells could implement a single logical
multilevel cell with $n!$ levels, where increasing the logical cell's
level by $1$ corresponds to a single transition in the $n$-RMGC. This
allows for a natural integration of rank modulation with other
multilevel approaches such as rewriting schemes
\cite{YaaVarSieWol08,JiaLanSchBru09,ChiFinLiuMit10,JiaBohBru10}.

Other recent works have explored error-correcting codes for rank
modulation, where different types of errors are addressed by a careful
choice of metric. In \cite{JiaSchBru10}, Kendall's $\tau$-metric was
considered, since a small charge-constrained error translates into a
small distance in the metric. In contrast, the $\ell_\infty$-metric
was used in \cite{TamSch10,KloLinTsaTze10}, as small distances in the
metric correspond to small limited-magnitude errors.

In this paper, we explore Gray codes for rank modulation which detect
a single error, under both metrics mentioned above. Such codes are
known as \emph{snake-in-the-box codes}, and have been studied
extensively for binary vectors with the Hamming metric and with
single-bit flips as allowable transitions (see \cite{AbbKat91} and
references therein).

The paper is organized as follows: In Section \ref{preliminaries} we
present basic notation and definitions. In Section \ref{K-t_metric} we
review properties of Kendall's $\tau$-metric, present a recursive
construction of snake-in-the-box codes over the alternating groups of
odd orders, with asymptotically-optimal rate, then present auxiliary
functions needed for the use of codes generated by this construction,
and conclude by presenting upper-bounds on the size of such
snake-in-the-box codes. In Section \ref{ell_metric} we present a
direct construction of snake-in-the-box codes of every order in the
$\ell_\infty$-metric based on results from \cite{JiaMatSchBru09} which
we show have asymptotically-optimal rate, and also present some
required auxiliary functions. We conclude in Section \ref{conclusion}
with some ad-hoc results, as well as some open questions.

\section{Preliminaries}\label{preliminaries}

We shall denote by $\sigma=[ a_1, a_2 , \ldots , a_n ]$ the
permutation over $[n]\triangleq\mathset{1,2,\ldots,n}$ such that for
all $i\in[n]$ it holds that $\sigma(i)=a_i$ (and, naturally,
$\mathset{a_1,a_2,\ldots,a_n}=[n]$). This form is called the
\emph{vector notation} for permutation. We let $S_{n} =
\mathrm{Sym}[n]$ be the symmetric group on $[n]$, and $A_{n}\leq
S_{n}$ be the alternating group of the same order. For $\sigma,\tau\in
S_n$, their composition, denoted $\sigma\tau$, is the permutation for
which $\sigma\tau(i)=\sigma(\tau(i))$ for all $i\in[n]$.

A cycle, denoted $(a_1,a_2,\dots,a_k)$, is a permutation mapping
$a_i\mapsto a_{i+1}$ for all $i\in[k-1]$, as well as $a_k\mapsto a_1$.
We shall occasionally use \emph{cycle notation} in which a permutation
is described as a composition of cycles. We also recall that any
permutation may be represented as a composition of cycles of size $2$,
and that the parity of the number of these cycles does not depend on
the decomposition. Thus we have \emph{even} and \emph{odd}
permutations, with positive and negative \emph{signs}, respectively.

\begin{definition}
Given a set $S$ and a subset of transformations $T\subseteq S^{S} =
\mathsetp{f}{f:S\rightarrow S}$, a \emph{Gray code} over $S$, using
transitions $T$, of size $M\in\N$, is a sequence
$C=(c_0,c_1,\dots,c_{M-1})$ of $M$ distinct elements of $S$, called
\emph{codewords}, such that for all $j\in[M-1]$ there exists $t\in T$
such that $c_{j}=t(c_{j-1})$.
\end{definition}

Alternatively, when the original permutation $c_{0}$ is known (or
irrelevant), we use a slight abuse of notation in referring to the
sequence of transformations $(t_{k_1},\ldots,t_{k_{M-1}})$
generating the code (i.e., $c_{j}=t_{k_j}(c_{j-1})$) as the code
itself.

In the above definition, when $M=\abs{S}$ the Gray code is called
\emph{complete}. If there exists $t\in T$ such that
$t\parenv{c_{M-1}}=c_{0}$ the Gray code is called \emph{cyclic}, $M$
is called its \emph{period}, and we shall, when listing the code by
its sequence of transformations, include $t_{k_M}\triangleq t$ at the
end of the list. The \emph{rate} of $C$, denoted $R(C)$, is defined as
\[R(C)\triangleq\frac{\log_2 M}{\log_2 \abs{S}}.\]

In the context of rank modulation for flash memories, the set of
transformations $T$ comprises of \pttt operations, first
used in \cite{JiaMatSchBru09}, and later also in
\cite{Sch10,WanBru10,EngLanSchBru10}. We denote by
$t_{i}\in\mathrm{Aut}\parenv{S_{n}}$ the \quo{push-to-the-top}
operation on index $i$, i.e.,
\begin{multline*}
t_i [a_1, a_2, \ldots, a_{i-1}, a_i, a_{i+1}, \ldots, a_n ] = \\
=[ a_i, a_1, a_2, \ldots, a_{i-1}, a_{i+1}, \ldots, a_n],
\end{multline*}
and throughout the paper set
$T=\mathset{t_2,t_3,\dots,t_n}$. Restricting the transformations to
\pttt operations allows fast cell programming, and
eliminates overshoots (see \cite{JiaMatSchBru09}).

For ease of presentation only, we also denote by $\underline{t}_i$ the
\quo{push-to-the-bottom} operation on index $n+1-i$, i.e.,
\begin{multline*}
\underline{t}_i [a_1, a_2, \ldots, a_{n-i}, a_{n+1-i}, a_{n+2-i}, \ldots, a_n ] = \\
=[ a_1, a_2, \ldots, a_{n-i}, a_{n+2-i}, \ldots, a_n, a_{n+1-i}].
\end{multline*}

Let $d:S\times S\rightarrow\N\cup\mathset{0}$ be a distance function
inducing a metric $\cM$ over $S$. Given a transmitted codeword $c\in
C$ and its received version $\tilde{c}\in S$, we say a single error
occurred if $d(c,\tilde{c})=1$. We are interested in Gray codes
capable of detecting single errors, which we now define.

\begin{definition}
Let $\cM$ be a metric over $S$ induced by a distance measure $d$.  A
\emph{snake-in-the-box code} over $\cM$ and $S$, using transitions
$T$, is a Gray code $C$ also over $S$ and using $T$, in which for
every pair of distinct elements $c,c'\in C$, $c\neq c'$, one has
$d\parenv{c,c'}\geq 2$.
\end{definition}

Since throughout the paper, our ambient space is $S_n$, and the
transformations we use are the \pttt operations $T$, we shall
abbreviate our notation and call the snake-in-the-box code of size $M$
an \emph{$(n,M,\cM)$-snake}, or an \emph{$\cM$-snake}. We will be
considering two metrics in the next sections: Kendall's $\tau$-metric,
$\cK$, and the $\ell_\infty$-metric, with their respective
$\cK$-snakes and $\ell_\infty$-snakes.

It is interesting to note that the classical definition of
snake-in-the-box codes (see the survey \cite{AbbKat91}) is slightly weaker
in the sense that $d(c,c')\geq 2$ is required for distinct $c,c'\in
C$, \emph{unless} $c$ and $c'$ are adjacent in $C$. This, however, is
a compromise due to the fact that in the classical codes over binary
vectors, the transformations (which flip a single bit) always create
adjacent codewords at distance $1$ apart. This compromise is
unnecessary in our case since, as we shall later see, the
\pttt operations allow adjacent words at distance $2$ or
more apart.

\section{Kendall's $\tau$-Metric and $\cK$-Snakes}\label{K-t_metric}

Kendall's $\tau$-metric \cite{KenGib90}, denoted $\cK$, is induced by
the bubble-sort distance which measures the minimal amount of adjacent
transpositions required to transform one permutation into the
other. For example, the distance between the permutations
$\sparenv{2,1,4,3}$ and $\sparenv{2,4,3,1}$ is $2$, as
\[\sparenv{2,1,4,3} \rightarrow \sparenv{2,4,1,3} \rightarrow \sparenv{2,4,3,1}\]
is a shortest sequence of adjacent transpositions between the
two. More formally, for $\alpha,\beta\in S_{n}$, as noted in
\cite{JiaSchBru10},
\[d_{K}(\alpha,\beta) = \mathsetp{(i,j)}{\alpha(i)<\alpha(j)\wedge\beta(i)>\beta(j)}.\]
The metric $\cK$ was first introduced by Kendall \cite{KenGib90} in the
study of ranking in statistics. It was observed in \cite{JiaSchBru10} that
a bounded distance in Kendall's $\tau$-metric models errors caused by
bounded changes in charge-levels of cells in the flash
memory. Error-correcting codes for this metric were studied in
\cite{JiaSchBru10,BarMaz10}.

We let Kendall's $\tau$ \emph{adjacency graph} of order $n\in\N$ be
the graph $G_{n}=\parenv{V_{n},E_{n}}$ whose vertices are the elements
of the symmetric group $V_{n}=S_{n}$, and $\bracenv{\alpha,\beta}\in
E_{n}$ if and only if $d_{K}(\alpha,\beta)=1$. It is well known that
Kendall's $\tau$-metric is \emph{graphic} \cite{DezHua98}, i.e., for
every $\alpha,\beta\in S_{n}$, $d_{K}(\alpha,\beta)$ equals the length
of the shortest path between the two in the adjacency graph, $G_n$.

\subsection{Construction}

We begin the construction process by restricting ourselves to Gray
codes using only \pttt operations on odd indices. The following lemma
provides the motivation for this restriction.

\begin{lemma}\label{only_even}
A Gray code over $S_n$ using only \pttt operations
on odd indices is a $\cK$-snake.
\end{lemma}
\begin{IEEEproof}
One can readily verify that a \pttt operation on an odd
index is an even permutation. Thus, the codewords in a Gray code using
only such operation are all with the same sign.

On the other hand, an adjacent transposition is an odd permutation, thus,
flipping the sign of the permutation it acts on. It follows that in a list
of codewords, all with the same sign, there are no two codewords which
are adjacent in $G_n$, i.e., the Gray code is a $\cK$-snake.
\end{IEEEproof}

Lemma \ref{only_even} saves us the need to check whether a Gray code
is in fact a $\cK$-snake, at the cost of restricting the set of
allowed transitions. In particular, if $n$ is even, the last element
cannot be moved. By starting with an even permutation and using only
\pttt operations on odd indices we get a sequence of
even permutations, i.e., from the alternating group of same order.
Thus, throughout this part, the context of the alternating group
$A_{2n+1}$ is assumed, where $n\in\N$.

The construction we are about to present is recursive in nature.  As a
base for the recursion, we note that three consecutive
\quo{push-to-the-top} operations on the 3rd index of permutations in
$A_3$ constitute a complete cyclic $(3,3,\cK)$-snake:
\[C_3\triangleq \parenv{[1,2,3],[3,1,2],[2,3,1]}.\]

Now, assume that there exists a cyclic $(2n-1,M_{2n-1},\cK)$-snake,
$C_{2n-1}$, and let
\[t_{k_1},t_{k_2},\dots,t_{k_{M_{2n-1}}}\]
be the sequence of transformations generating it, where $k_j$ is odd
for all $j\in[M_{2n-1}]$. We also assume that $k_1=2n-1$ (this
requirement, while perhaps appearing arbitrary, is actually quite
easily satisfied. Indeed, every sufficiently large cyclic $\cK$-snake
over $S_{2n-1}$ must, WLOG, satisfy it. We shall make it a point to
demonstrate that this holds for our construction).

We fix arbitrary values for $a_0,a_1,\dots,a_{2n-2}$ such that
\begin{equation}
\label{eq:fixa}
\mathset{a_0,a_1,\ldots,a_{2n-2}}=[2n+1]\setminus\mathset{1,3}.
\end{equation}
Throughout the paper we shall take the indices of $a$ to be modulo
$2n-1$. For all $i\in[2n-1]$ we define
\[\sigma^{(i)}_0\triangleq[1, a_{i}, 3, a_{i+1}, \ldots, a_{i+2n-2}],\]
such that we indeed have $\sigma^{(i)}_0\in A_{2n+1}$, i.e.,
$\sigma^{(i)}_0$ is an even permutation (one simple way of achieving
this is to choose them in ascending order).

We now define for all $i\in[2n-1]$ and $j\in[M_{2n-1}]$ the permutation 
\[\sigma^{(i)}_{j(2n+1)}\triangleq\underline{t}_{k_{j}}\parenv{\sigma^{(i)}_{(j-1)(2n+1)}},\]
i.e., we construct cycles corresponding to a mirror view of $C_{2n-1}$
on all but the two uppermost indices of $\sigma^{(i)}_0$ (which, as we
recall, are $\parenv{1, a_{i}}$). We now note the following properties
of our construction:

\begin{lemma}\label{K-t_mid_construction}
Let $i,k\in[2n-1]$ and $j,l\in[M_{2n-1}]$. The following are equivalent:
\begin{enumerate}
\item 
  The permutations $\sigma^{(i)}_{j(2n+1)}$ and $\sigma^{(k)}_{l(2n+1)}$
  are cyclic shifts of each other.
\item 
  $\sigma^{(i)}_{j(2n+1)} = \sigma^{(k)}_{l(2n+1)}$.
\item 
  $i=k$ and $j=l$.
\end{enumerate}
\end{lemma}
\begin{IEEEproof}
First, if $\sigma^{(i)}_{j(2n+1)}$ is a cyclic shift of
$\sigma^{(k)}_{l(2n+1)}$, since
\[\sigma^{(i)}_{j(2n+1)}(1) = 1 = \sigma^{(k)}_{l(2n+1)}(1)\]
then necessarily
\[\sigma^{(i)}_{j(2n+1)} = \sigma^{(k)}_{l(2n+1)}.\]
It then follows that
\[a_i = \sigma^{(i)}_{j(2n+1)}(2) = \sigma^{(k)}_{l(2n+1)}(2) = a_k,\]
hence $i=k$. Moreover, since the two permutations' last $n-1$ elements
agree, and $t_{k_1},t_{k_2},\dots,t_{k_{M_{2n-1}}}$ induce a Gray
code, then $j=l$.

Finally, that the last statement implies the first is trivial.
\end{IEEEproof}

\begin{lemma}\label{K-t_mid_cyclic}
For all $i\in[2n-1]$ it holds that
\[\sigma^{(i)}_{M_{2n-1}(2n+1)}=\sigma^{(i)}_0.\]
\end{lemma}
\begin{IEEEproof}
The transformations $t_{k_1},t_{k_2},\dots,t_{k_{M_{2n-1}}}$ induce a cyclic code,
and the claim follows directly.
\end{IEEEproof}

Therefore we have constructed $2n-1$ cycles comprised of cyclically
non-equivalent permutations (although, at this point they are not
generated by \quo{push-to-the-top} operations).

It shall now be noted that 
\[\underline{t}_k = t_{2n+1}^{2n}t_{2n+2-k}.\]
Hence, if we define for all $i\in[2n-1]$, $0\leq j<M_{2n-1}$, and
$1<m\leq 2n$, the permutations
\begin{align*}
\sigma^{(i)}_{j(2n+1)+1} & \triangleq
t_{2n+2-k_{j+1}}\sigma^{(i)}_{j(2n+1)}\\
\sigma^{(i)} _{j(2n+1)+m} & \triangleq t_{2n+1}^{m-1}\sigma^{(i)}_{j(2n+1)+1},
\end{align*}
then it holds that
\[\sigma^{(i)}_{(j+1)(2n+1)}=t_{2n+1}\sigma^{(i)}_{j(2n+1)+2n}.\]

Our observation from one paragraph above means that at this point we
have $2n-1$ disjoint cycles, which we conveniently denote
\[C^{(i)}_{2n+1}\triangleq\parenv{\sigma^{(i)}_0,\sigma^{(i)}_1,\dots,
\sigma^{(i)}_{M_{2n-1}(2n+1)-1}},\]
for all $i\in[2n-1]$ (for ease of notation, we let $C^{(0)}_{2n+1} = 
C^{(2n-1)}_{2n+1}$). Each of the cycles is of size $(2n+1)M_{2n-1}$, 
is generated by \quo{push-to-the-top} operations, and contains all cyclic
shifts of elements present in our previous version of that cycle.

\begin{figure}[t]
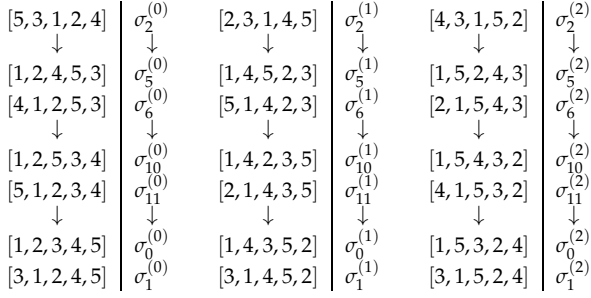

\begin{footnotesize}
\[
\begin{array}{c|c}
\sparenv{5,3,1,2,4} & \sigma_2^{(0)}\\
\downarrow          & \downarrow \\
\sparenv{1,2,4,5,3} & \sigma_5^{(0)}\\
\sparenv{4,1,2,5,3} & \sigma_6^{(0)}\\
\downarrow          & \downarrow \\
\sparenv{1,2,5,3,4} & \sigma_{10}^{(0)}\\
\sparenv{5,1,2,3,4} & \sigma_{11}^{(0)}\\
\downarrow          & \downarrow \\
\sparenv{1,2,3,4,5} & \sigma_0^{(0)}\\
\sparenv{3,1,2,4,5} & \sigma_1^{(0)}
\end{array}
\quad
\begin{array}{c|c}
\sparenv{2,3,1,4,5} & \sigma_2^{(1)}\\
\downarrow          & \downarrow \\
\sparenv{1,4,5,2,3} & \sigma_5^{(1)}\\
\sparenv{5,1,4,2,3} & \sigma_6^{(1)}\\
\downarrow          & \downarrow \\
\sparenv{1,4,2,3,5} & \sigma_{10}^{(1)}\\
\sparenv{2,1,4,3,5} & \sigma_{11}^{(1)}\\
\downarrow          & \downarrow \\
\sparenv{1,4,3,5,2} & \sigma_0^{(1)}\\
\sparenv{3,1,4,5,2} & \sigma_1^{(1)}
\end{array}
\quad
\begin{array}{c|c}
\sparenv{4,3,1,5,2} & \sigma_2^{(2)}\\
\downarrow          & \downarrow \\
\sparenv{1,5,2,4,3} & \sigma_5^{(2)}\\
\sparenv{2,1,5,4,3} & \sigma_6^{(2)}\\
\downarrow          & \downarrow \\
\sparenv{1,5,4,3,2} & \sigma_{10}^{(2)}\\
\sparenv{4,1,5,3,2} & \sigma_{11}^{(2)}\\
\downarrow          & \downarrow \\
\sparenv{1,5,3,2,4} & \sigma_0^{(2)}\\
\sparenv{3,1,5,2,4} & \sigma_1^{(2)}
\end{array}
\]
\end{footnotesize}
\caption{ A $(5,45,\cK)$-snake, $C_5$, from Theorem
  \ref{ksnakes}. Down arrows stand for an omitted sequence of $t_5$
  transformations. The transition from column to column uses a single
  $t_3$ transformation. }
\label{ex:ksnake}
\end{figure}

\begin{theorem} \label{ksnakes}
Given a cyclic $(2n-1,M_{2n-1},\cK)$-snake using only \pttt
operations on odd indices, and such that its first
transformation is $t_{2n-1}$, there exists a cyclic
$(2n+1,M_{2n+1},\cK)$-snake with the same properties, whose size is
$M_{2n+1}=(2n-1)(2n+1)M_{2n-1}$.
\end{theorem}
\begin{IEEEproof}
Since $k_1=2n-1$, it holds for all $i\in[2n-1]$ that $\sigma^{(i)}_1=t_3\sigma^{(i)}_0$, and we recall $\sigma^{(i)}_2 = t_{2n+1}\sigma^{(i)}_1$. More explicitly,
\begin{align*}
\sigma^{(i)}_1 & = \sparenv{3, 1, a_{i}, a_{i+1}, \ldots, a_{i+2n-2}}\\
\sigma^{(i)}_2 & = \sparenv{a_{i+2n-2}, 3, 1, a_{i}, a_{i+1}, \ldots, a_{i+2n-3}},
\end{align*}
where, again, the indices are taken modulo $2n-1$. Thus for all
$i\in[2n-2]$ we have
\[t_3\sigma^{(i)}_1 = \sparenv{a_{i}, 3, 1, a_{i+1}, \ldots, a_{i+2n-2}} = \sigma^{(i+1)}_2\] 
and $t_3\sigma^{(2n-1)}_1 = \sigma^{(1)}_2$.

Let $E$ denote the left-shift operator, and so
\[E^2 C^{(i)}_{2n+1}=\parenv{\sigma^{(i)}_2,\sigma^{(i)}_3,\dots,
\sigma^{(i)}_{M_{2n-1}(2n+1)-1},\sigma^{(i)}_0,\sigma^{(i)}_1}.\]
By the above observations we conclude that
\[C_{2n+1}\triangleq E^2 C^{(0)}_{2n+1},E^2 C^{(1)}_{2n+1},\dots,
E^2 C^{(2n-2)}_{2n+1}\] is a cyclic $(2n+1,M_{2n+1},\cK)$-snake,
consisting of
\[M_{2n+1}=(2n-1)(2n+1)M_{2n-1}\]
permutations. The code $C_{2n+1}$ obviously uses $t_{2n+1}$, and so
some cyclic shift of it has it as its first transition (in fact, for
every $i\in[2n-1]$ one has $\sigma^{(i)}_{3} =
t_{2n+1}\sigma^{(i)}_{2}$, and in particular, $E^{2}C^{(0)}_{2n+1}$
has $t_{2n+1}$ as its first transition, and so does
$C_{2n+1}$). Finally, it is easily verifiable that all \pttt
operations are on odd indices. (See an example in Figure
\ref{ex:ksnake}.)
\end{IEEEproof}

A property of rank-modulation cell programming is that an erasure of
an entire cell block is required only when a specific cell is to
exceed its maximal permitted charge level. It is therefore of interest
to analyze the rate with which our constructed codes increase the
charge level of any given cell.

Repeated \pttt operations on a given cell will result in a fast
increase in that cell's charge level, and growing gaps between it and
the charge levels of other cells. It is therefore most cost-economic,
in the sense that it delays the need for a time-consuming erasure and
reprogramming cycle, to employ a programming strategy which retains
the charge levels of individual cells as balanced as possible. Such
balanced Gray codes were constructed in \cite{JiaMatSchBru09}.

In this part's context, this goal is achieved if and only if every two
subsequent incidents in a cyclic $\parenv{2n+1,M,\cK}$-snake where a
\pttt operation is applied to a certain cell are separated by at most
$2n+1$ operations on other cells.  Our family of codes nearly achieves
this goal:

\begin{lemma}\label{K-t_balancing}
For every permutation $\sigma\in C_{2n+1}$, in the $\cK$-snake
constructed in Theorem \ref{ksnakes}, there exists another
$\sigma^{\prime}\in C_{2n+1}$ such that
$\sigma(1)=\sigma^{\prime}(1)$, following it by no more than $2n+3$
steps.
\end{lemma}
\begin{IEEEproof}
Recall that
\[C_{2n+1} = E^2 C^{(0)}_{2n+1},E^2 C^{(1)}_{2n+1},\dots,E^2 C^{(2n-2)}_{2n+1}.\]
By the nature of our construction, for $n\geq 2$, every \pttt
operation, on all but the last rank in the code, appears either as
part of the pattern
\[\dots,\underbrace{t_{2n+1},\dots,t_{2n+1}}_{2n},t_i,\underbrace{t_{2n+1},\dots,t_{2n+1}}_{2n},\dots\]
or as
\[\dots,\underbrace{t_{2n+1},\dots,t_{2n+1}}_{2n},t_3,t_3,\underbrace{t_{2n+1},\dots,t_{2n+1}}_{2n},\dots\]
It is therefore the case that there exist $0\leq k\leq 2n$ and
$j\in[n]$ such that the transformations used in $C_{2n+1}$ after
$\sigma$ are of the following two forms:
\begin{enumerate}
\item $\underbrace{t_{2n+1},\dots,t_{2n+1}}_{k},t_{2j+1},\underbrace{t_{2n+1},\dots,t_{2n+1}}_{2n}$
\item $\underbrace{t_{2n+1},\dots,t_{2n+1}}_{k},t_3,t_3,\underbrace{t_{2n+1},\dots,t_{2n+1}}_{2n}$
\end{enumerate}
In the second case, one notes:
\[\sigma(1) = \begin{cases}
t_{2n+1}^{2n-1} t_3^2\sigma(1) & k=0\\
t_3^2 t_{2n+1}\sigma(1) & k=1\\
t_3 t_{2n+1}^{2}\sigma(1) & k=2\\
t_{2n+1}^{2n+1-k} t_3^2 t_{2n+1}^{k}\sigma(1) & k>2.\end{cases}\]
Finally, in the first case, we note that
\[\sigma(1) = \begin{cases}
t_{2n+1}^{2n-k} t_{2j+1} t_{2n+1}^{k}\sigma(1) & k<2j+1\\
t_{2j+1} t_{2n+1}^{k}\sigma(1) & k=2j+1\\
t_{2n+1}^{2n+1-k} t_{2j+1} t_{2n+1}^{k}\sigma(1) & k>2j+1.\end{cases}\]
\end{IEEEproof}

It is of interest to note that, of all cases discussed in the last proof, 
the second case where $k>2$ is the only situation in which 
another instance of programming to the specific cell fails to occur in 
$2n+2$ steps, i.e., for the large majority of cases (in all but 
$\frac{2n-1}{M_{2n+1}}$ of them), the construction of 
Theorem \ref{ksnakes} yields optimally-behaving codes.

We now turn to consider the rate of the constructed codes, and show that
it is asymptotically optimal.

\begin{theorem}
The $\cK$-snakes constructed in Theorem \ref{ksnakes} have an
asymptotically-optimal rate.
\end{theorem}
\begin{IEEEproof}
Starting from our base case of a complete cyclic $(3,3,\cK)$-snake, we
define for all $n\in\N$ the ratio
\[D_{2n+1}\triangleq\frac{M_{2n+1}}{(2n+1)!},\]
which is the size of
our constructed code over the total size of $S_{2n+1}$.  We note that
\[\frac{D_{2n+1}}{D_{2n-1}} = \frac{M_{2n+1}\cdot
  (2n-1)!}{(2n+1)!\cdot M_{2n-1}} = \frac{2n-1}{2n}.\]
Therefore,
since $D_3=\frac{1}{2}$, we have for all $2\leq n\in\N$ that
\[D_{2n+1} = \frac{1}{2}\prod_{m=2}^{n}\frac{2m-1}{2m} = \frac{(2n)!}{n!^{2}\cdot 2^{2n}}.\]
Using Stirling's approximation one observes
\begin{align*}
\lim_{n\rightarrow\infty}D_{2n+1}\sqrt{\pi n} &= 
\lim_{n\rightarrow\infty}\frac{(2n)!\sqrt{\pi n}}{n!^{2}\cdot 2^{2n}}\\
&= 
\lim_{n\rightarrow\infty}\frac{\sqrt{4\pi n}\parenv{\frac{2n}{e}}^{2n}\sqrt{\pi n}}
{\parenv{\sqrt{2\pi n}\parenv{\frac{n}{e}}^n}^2\cdot 2^{2n}} = 1,
\end{align*}
and therefore it holds that
\[\lim_{n\rightarrow\infty}R(C_{2n+1})=\lim_{n\rightarrow\infty}\frac{\log_2 M_{2n+1}}{\log_2\abs{S_{2n+1}}} = 1.\]
\end{IEEEproof}

\subsection{Successor Calculation and Ranking Algorithms}

We now turn to present algorithms associated with the codes we
constructed in the previous section. The algorithms are brought here
for completeness of presentation, and are straightforward derivations
from the construction. We shall, therefore, only provide an intuitive
sketch of correctness for them, as we shall later do in the section
corresponding to $\ell_\infty$-snakes.

In order to use the codes described in Theorem \ref{ksnakes} in the
implementation of a logic cell (with $M_{2n+1}$ levels), importance is
known to the ability of efficiently increasing the cell's level, i.e.,
one needs to know for every given permutation in the code the
appropriate \pttt operation required to produce the subsequent permutation.

For the code $C_{2n+1}$ from Theorem \ref{ksnakes}, the function
$\ksucc\parenv{n,\sparenv{b_{1},\ldots,b_{2n+1}}}$ takes as input a
permutation in the code, and returns as output the index $i$ of the
required transformation $t_{i}$.  It is assumed throughout this part
that the elements $\mathset{a_{i}}_{i=0}^{2n-2}$ from \eqref{eq:fixa},
used in our construction, are known, and we will denote them with
superscript $(n)$ to indicate order when it is not clear from
context.  Furthermore, we require a function
\[\ind_n(b):[2n+1]\setminus\mathset{1,3}\rightarrow [0,2n-2]\]
which returns the unique index such that $a_{\ind_n(b)}=b$. We assume
$\ind_n$ runes in $O(1)$ time\footnote{%
  Though the integers used
  throughout are of magnitude $O(n)$, and so may require $O(\log n)$
  bits to represent, we tacitly assume (as in \cite{JiaMatSchBru09})
  all simple integer operations, e.g., assignment, comparison,
  addition, etc., to take $O(1)$ time. }. One possible way, among
many, of achieving this is by defining:
\[
a^{(n)}_i \triangleq \begin{cases}
2 & i=0 \\
i+3 & i\geq 1
\end{cases}
\qquad
\ind_n(b)\triangleq\begin{cases}
0 & b=2, \\
b-3 & b\geq 4.
\end{cases}
\]
Finally, we naturally assume validity of
the input in all procedures.

Our strategy will be to identify the vertices in $C_{2n+1}$ which
require a transformation other than $t_{2n+1}$. Those are either
permutations with leading $1$'s (those on which we initially performed
\quo{push-to-the-bottom} operations in our construction), or the last
permutation in each $E^2 C^{(j)}_{2n+1}$.  In the latter case we need
only apply $t_3$, where the former requires translation of the
$a^{(n)}_{i}$'s according to their respective positions in the
originating permutation of each $C^{(j)}_{2n+1}$, and a recursive run
of $\ksucc$ to determine the correct \quo{push-to-the-bottom}
operation to be performed.

It shall be noted at this point that a degree of freedom exists in the
cyclic shift of $C_{2n-1}$ one applies to construct each
$C_{2n+1}^{(j)}$ (one only needs to confirm that the first \pttt
operation shall be on the last index). This shift shall be denoted by
the following bijection for every order $n\in\N$ and index
$j\in[2n-1]$:
\[\dw{n}{j}:\mathset{3}\cup\mathset{a^{(n)}_{i}}_{i\neq j}\longrightarrow [2n-1],\]
defined such that the \quo{push-to-the-bottom} operation applied to 
\[\sparenv{1,a^{(n)}_{j},b_1,\ldots,b_{2n-1}}\in C^{(j)}_{2n+1}\]
matches the \pttt operation applied in $C_{2n-1}$ to 
\[\sparenv{\dw{n}{j}b_{2n-1},\dw{n}{j}b_{2n-2},\ldots,\dw{n}{j}b_{1}}.\]
We shall further denote its inverse as $\up{n}{j}{}$. These two
bijections can be implemented in $O(1)$ time, for example, by taking 
as a starting point $C_{2n-1}$'s $(2n-4)$-ranked permutation 
\[\sparenv{a^{(n-1)}_0,\ldots,a^{(n-1)}_{2n-4},3,1},\]
and defining accordingly
\begin{equation}
\label{eq:downtranslation}
\dw{n}{j} b = 
\begin{cases}
1 & b=3\\
3 & \ind_n(b)= j+1\\
a^{(n-1)}_{(j-\ind_n(b)-1)\bmod (2n-1)} & \text{otherwise},

\end{cases}
\end{equation}
where $\ind_n(b)=j+1$ is checked modulo $2n-1$, as well as
\begin{equation}
\label{eq:uptranslation}
\up{n}{j} b = 
\begin{cases}
3 & b=1\\
a^{(n)}_{j+1} & b=3\\
a^{(n)}_{j-\ind_{n-1}(b)-1} & \text{otherwise}.
\end{cases}
\end{equation}

\begin{function}[t]
\begin{footnotesize}
\DontPrintSemicolon
\Input{$n\in\N$, A permutation $[b_{1},\ldots,b_{2n+1}]\in C_{2n+1}$}
\Output{An odd $i \in \lbrace 3,\ldots,2n+1\rbrace$ that determines the transition $t_i$ to the next permutation in $C_{2n+1}$}
	\If {$n=1$}
	{
		\Return $3$ \nllabel{alg:K-t_successor:base}
	}
	\If {$b_{1}=3$ and $b_{2}=1$ and $\forall 3\leq i\leq 2n:$ $(\ind_n\parenv{b_{i+1}} - \ind_n\parenv{b_{i}})\equiv 1 \pmod{2n-1}$}
	{
		\Return $3$ \nllabel{alg:K-t_successor:cycle_shift}
	}
	\If {$b_{1}=1$}
	{
		$j\leftarrow \ind_n\parenv{b_2}$\\
		$i\leftarrow \ksucc\parenv{n-1,\sparenv{\dw{n}{j}b_{2n+1},\dw{n}{j}b_{2n},\ldots,\dw{n}{j}b_{3}}}$\\
		\Return $2n+2-i$ \nllabel{alg:K-t_successor:in_cycle}\\
	}
	\Return $2n+1$ \nllabel{alg:K-t_successor:shift}\\
\end{footnotesize}
\caption{() $\ksucc\parenv{n,\sparenv{b_1,\ldots,b_{2n+1}}}$}
\label{alg:K-t_successor}
\end{function}

\begin{lemma}\label{K-t_succ_complx}
$\ksucc$ runs in $O(1)$ amortized time.
\end{lemma}
\begin{IEEEproof}
We first note that by the nature of our construction the element $1$
appears in the leading index precisely $(2n-1)\cdot M_{2n-1}$ times,
which constitutes $\frac{1}{2n+1}$ of the code's size. The pair
$(3,1)$ leads no more (and in fact strictly less) permutations.

Therefore, if we let $E_{n}$ denote the expected number of steps
performed by $\ksucc$ when called on input of length $2n+1$, then we
note the recursive connection
\begin{align*}
E_{n} & \leq O(1) + \frac{1}{2n+1}O(n) + \frac{1}{2n+1}\parenv{O(n)+E_{n-1}} \\
 & = O(1) + \frac{1}{2n+1}E_{n-1}.
\end{align*}
Developing this inequality recursively, there exists $L\in\N$ such that
\begin{align*}
E_{n} \leq & L+\frac{1}{2n-1}E_{n-1}\\
\leq & \parenv{1+\frac{1}{2n-1}}L + \frac{1}{(2n-1)(2n-3)}E_{n-2} \leq \\
\vdots \\
\leq & \parenv{1+\frac{1}{2n-1}+\frac{n-2}{(2n-1)(2n-3)}}L + \frac{n!2^{n}}{(2n)!}E_{1},
\end{align*}
and so $E_n=O(1)$.
\end{IEEEproof}

To use $C_{2n+1}$ in the implementation of a logic cell, one also
needs a method of computing a given permutation's rank in the code. We
implement the function $\krank\parenv{[b_{1},\ldots,b_{2n+1}]}$ which
receives as input a permutation $[b_{1},\ldots,b_{2n+1}]\in C_{2n+1}$
and returns its rank in
\[C_{2n+1} = E^2 C^{(0)}_{2n+1},E^2 C^{(1)}_{2n+1},\dots,E^2 C^{(2n-2)}_{2n+1},\]
in the order indicated by that notation. The assumptions made in the
previous part are still in effect. Moreover, we will require knowledge
of the cyclic shift of $C_{2n-1}$ used in the construction of each
$C^{(j)}_{2n+1}$, which we retain in the form of $r^{(j)}_{2n+1}$, the
rank of permutation in $C_{2n-1}$ which was chosen as a starting
point.  For example, in the method suggested by
\eqref{eq:downtranslation} and \eqref{eq:uptranslation}, we have
\[r^{(j)}_{2n+1} = 2n-4\]
for all $j\in[2n-1]$.

We use the following method: first identify the position of $1$ in the
permutation, and the following element, which gives us both the
subcode the permutation belongs to and the cyclic shift in our mock
\quo{push-to-the-bottom} operation.  Armed with that information we
then scan the permutation backwards and translate the $a^{(n)}_j$'s
indices according to the subcode in the same way we did in $\ksucc$.
After that, a recursive run of $\krank$ will give us the permutation's
position in its subcode, which we will combine with the cyclic shift
to produce the correct rank, taking $r^{(j)}_{2n+1}$ into account 
and remembering that $C_{2n+1}$ is constructed of the 
$E^{2}C^{(j)}_{2n+1}$'s rather than the $C^{(j)}_{2n+1}$'s.

\begin{function}[t]
\begin{footnotesize}
\DontPrintSemicolon
\Input{A permutation $[b_{1},\ldots,b_{2n+1}]\in C_{2n+1}$}
\Output{The rank $k\in\mathset{ 0,\ldots,M_{2n+1}-1}$ associated with the given permutation in $C_{2n+1}$}
	\If {$n=1$}
	{
		\Return $3-b_{2}$ \nllabel{alg:K-t_rank:base}\\
	}
	$i\leftarrow \min\mathsetp{l\in[2n+1]}{b_l=1}$\\
	$j\leftarrow \ind_n\parenv{b_{(i\bmod (2n+1))+1}}$\\
	\For {$l\leftarrow 1$ \KwTo $2n-1$}
	{
		$c_l\leftarrow \dw{n}{j}b_{\parenv{(i-l-1)\bmod (2n+1)}+1}$
	}
	$\mvar{r}\leftarrow \parenv{\krank\parenv{\sparenv{c_{1},\ldots,c_{2n-1}}}-r^{(j)}_{2n+1}} \bmod M_{2n-1}$\\
	$\mvar{rn}\leftarrow \parenv{(2n+1)(\mvar{r}-1)-1+\parenv{(i-2)\bmod (2n+1) }}\bmod ((2n+1)M_{2n-1}) $\\
	\Return $(2n+1)M_{2n-1}\cdot j + \mvar{rn}$ \nllabel{alg:K-t_rank:recursive}
\end{footnotesize}
\caption{() $\krank\parenv{\sparenv{b_1,\ldots,b_{2n+1}}}$}
\label{alg:K-t_rank}
\end{function}

\begin{lemma}\label{K-t_rank_complx}
The function $\krank$ operates in $O(n^{2})$ steps.
\end{lemma}
\begin{IEEEproof}
We note that $\krank$ performs $O(n)$ operations before calling upon itself with an 
order reduced by one. It therefore operates in $O(n^{2})$ time.
\end{IEEEproof}

Unranking permutations, i.e., the process of assigning to a given rank
in $[0,M_{2n+1}-1]$ the corresponding permutation in the $C_{2n+1}$,
might also be needed if one requires the logic cell to perform as more
than a counter. We implement a function $\kunrank(n,k)$ which returns
as output the $k$-ranked permutation in $C_{2n+1}$.

Naturally, all assumptions made above still hold. We will follow the
same general method used for $\krank$, i.e., we shall compute
$j\in[2n-1]$ such that the given rank belongs to $\sigma\in
E^{2}C_{2n+1}^{(j)}$, then adjust the rank to indicate the correct
position in $C_{2n+1}^{(j)}$. It will then remain to compute the
correct permutation in the \quo{push-to-the-bottom} cycle using a
recursive run, and shift it the required number of times.

\begin{function}[t]
\begin{footnotesize}
\DontPrintSemicolon
\Input{$n\in\N$; rank $k\in[0,M_{2n+1}-1]$}
\Output{The permutation $[b_{1},\ldots,b_{2n+1}]$ which is $k$th in $C_{2n+1}$}
	\If {$n=0$}
	{
		\Return $[1]$ \nllabel{alg:K-t_unrank:base}\\
	}

	$j\leftarrow \floorenv{\frac{k}{(2n+1)\cdot M_{2n-1}}}$\\
	$\mvar{pos}\leftarrow k\bmod ((2n+1)M_{2n-1})$\\
	$\mvar{perm}\leftarrow \parenv{\floorenv{\frac{\mvar{pos}+1}{2n+1}}+1+r^{(j)}_{2n+1}}\bmod M_{2n-1}$\\
	$\mvar{shift}\leftarrow \parenv{\mvar{pos}+2}\bmod (2n+1) $\\
	$\sparenv{c_{1},\ldots,c_{2n-1}}\leftarrow\kunrank(n-1,\mvar{perm})$\\
	\Return $t_{2n+1}^{\mvar{shift}}\sparenv{1,a^{(n)}_{\mvar{j}},\up{n}{j}c_{2n-1},\up{n}{j}c_{2n-2},\ldots,\up{n}{j}c_{1}}$  \nllabel{alg:K-t_unrank:general}
\end{footnotesize}
\caption{() $\kunrank\parenv{n, k}$}
\label{alg:K-t_unrank}
\end{function}

\begin{lemma}\label{K-t_unrank_complx}
The function $\kunrank$ operates in $O(n^{2})$ steps as well.
\end{lemma}
\begin{IEEEproof}
Follows exactly the same lines as our proof to Lemma \ref{K-t_rank_complx}.
\end{IEEEproof}

\subsection{Bounds on $\cK$-Snakes}

We begin by noting a simple upper bound on the size of $\cK$-snakes.

\begin{lemma}\label{K-t_full_paving}
If $C$ is an $(n,M,\cK)$-snake then
\begin{enumerate}
\item
  $M\leq\frac{1}{2}\abs{S_{n}}$.
\item
  $M=\frac{1}{2}\abs{S_{n}}$ if and only if for all
  $\bracenv{\alpha,\beta}\in E_{n}$ it holds that $\alpha\in C$ or
  $\beta\in C$.
\end{enumerate}
\end{lemma}
\begin{IEEEproof}
Every $\alpha\in S_{n}$ has exactly $(n-1)$ neighbors in $G_n$. When
we sum the edges for every vertex in $G_n$, each edge in $E_{n}$ is
counted precisely twice, hence
\[\abs{E_{n}}=\frac{n-1}{2}\cdot\abs{S_{n}} = \frac{n!(n-1)}{2}.\]

On the other hand, for every $\alpha,\beta\in C$ and $e_{1},e_{2}\in
E_{n}$ such that $\alpha\in e_{1}$ and $\beta\in e_{2}$ clearly
$e_{1}\neq e_{2}$. It follows that there are no less than $M(n-1)$
distinct edges in $E_{n}$. Hence 
\[M\leq\frac{1}{2}\abs{S_{n}}.\]

Finally, we note that $M=\frac{1}{2}\abs{S_{n}}$ iff
$M(n-1)=\abs{E_{n}}$, iff every edge in $E_{n}$ contains a (unique)
element of $C$.
\end{IEEEproof}

The codes we constructed in the previous section use only \pttt
operations on odd indices. We would now like to show that using even a
single \pttt operation on an even index can never result in a code
attaining the bound of Lemma \ref{K-t_full_paving} with equality. We
first require a simple lemma.

\begin{lemma}\label{K-t_odd_paths}
Let $C$ be a $\cK$-snake over $S_{n}$. If $\sigma,\sigma'\in
C$ and there exists a path in $G_n$ of odd length between them, then
that path contains an edge both of whose endpoints are not in $C$.
\end{lemma}
\begin{IEEEproof}
Consider such a path of odd length in $G_n$, connecting $\sigma$ and
$\sigma'$. Now color the vertices of $C$ black, and those of
$S_n\setminus C$ white.  Since $C$ is a $\cK$-snake, no edge in $E_n$
has both its ends colored black.  In the path above the vertices
cannot alternate in color since $\sigma$ and $\sigma'$ are colored
black and the path has odd length. It follows that there is an edge in
the path with both ends colored white, as claimed.
\end{IEEEproof}

A direct result of this lemma is presented in the following theorem:

\begin{theorem}\label{K-t_odd_paving_prefix}
If an $(n,M,\cK)$-snake $C$ contains a \pttt operation
on an even index then $M<\frac{1}{2}\abs{S_{n}}$.
\end{theorem}
\begin{IEEEproof}
We note that a single adjacent transposition acting on a permutation
flips the permutation's sign. Furthermore, a \pttt operation $t_i\in
T$, is equivalent to a sequence of $i-1$ adjacent transpositions
moving the $i$th element of the permutation to the first
coordinate. Thus, \pttt operations on even indices flip the
permutation's sign, while those on odd indices preserve it.

It readily follows that $\sigma,\sigma'\in S_n$ have different signs
iff \emph{every} path connecting them in $G_n$ has odd length.  Now,
if $\sigma'=t_{2m}(\sigma)$ for some $2m\in[n]$, and both are in $C$,
then they differ in sign and so by Lemma \ref{K-t_full_paving}(b) and
Lemma \ref{K-t_odd_paths}, $M<\frac{1}{2}\abs{S_{n}}$.
\end{IEEEproof}

We now aim to show a tighter upper-bound on the size of $\cK$-snakes
employing a \pttt operation on an even index.

\begin{theorem}\label{K-t_odd_paving}
If an $(n,M,\cK)$-snake $C$ contains a \pttt operation
on an even index then
\[M \leq \frac{1}{2}\abs{S_{n}} - \frac{1}{n-1}\binom{\floorenv{n/2}-1}{2}.\]
\end{theorem}
\begin{IEEEproof}
Let $C=\parenv{\sigma_1,\dots,\sigma_M}$.  We take $i\in[M-1]$ such
that $\sigma_{i+1} = t_{2m}\parenv{\sigma_{i}}$, where $2m\in[n]$. For
all $k,l\in\sparenv{\floorenv{\frac{n}{2}}-1}$, $k<l$, we define
\[k'\triangleq \begin{cases}k & k<m\\ k+1 & k\geq m\end{cases}
\qquad
l'\triangleq \begin{cases}l & l<m\\ l+1 & l\geq m.\end{cases}\]
For each $k'$ and $l'$ we can now define the paths in $G_n$
\[\sigma_i\rightarrow \omega^{(k',l')}_{1}
\rightarrow \omega^{(k',l')}_{2}
\rightarrow \dots
\rightarrow \omega^{(k',l')}_{2m+2}
\rightarrow \sigma_{i+1}\]
in the following recursive manner:
\begin{align*}
\omega^{(k',l')}_{1} & \triangleq \sigma_{i}(2k'-1,2k')\\
\omega^{(k',l')}_{2} & \triangleq \omega^{(k',l')}_{1}(2l'-1,2l'),
\end{align*}
for all $j\in[2m-1]$ we define
\begin{align*}
\omega^{(k',l')}_{j+2} & \triangleq \omega^{(k',l')}_{j+1}\parenv{2m-j,2m-j+1},
\end{align*}
and finally
\begin{align*}
\omega^{(k',l')}_{2m+2} & \triangleq \omega^{(k',l')}_{2m+1}(2l'-1,2l')\\
\omega^{(k',l')}_{2m+3} & \triangleq \omega^{(k',l')}_{2m+2}(2k'-1,2k') = \sigma_{i+1}.
\end{align*}

We note that these $\binom{\floorenv{n/2}-1}{2}$ paths are all of size
$2m+3$, connecting $\sigma_{i}$ and $\sigma_{i+1}$. Moreover, they
only possibly ever intersect in the first or last two vertices. It
follows from Lemma \ref{K-t_odd_paths} that each contains an edge
disjoint from $C$, and since we know each path's first and last edge
does intersect $C$, there therefore exist at least
$\binom{\floorenv{n/2}-1}{2}$ distinct edges in $G_n$ disjoint from
$C$. We can now improve upon the upper-bound from Lemma
\ref{K-t_full_paving} in the following way:
\[M(n-1)\leq \frac{n!(n-1)}{2} - \binom{\floorenv{n/2}-1}{2}\]
and reordering gives us the claim.
\end{IEEEproof}

\section{The $\ell_{\infty}$-Metric and $\ell_\infty$-Snakes}\label{ell_metric}

The $\ell_{\infty}$-metric is induced on $S_{n}$ by the embedding in
$\Z^{n}$ implied by the vector notation. More precisely, for
$\alpha,\beta\in S_{n}$ one defines
\[d_{\infty}(\alpha,\beta) = \max_{i\in[n]} \abs{\alpha(i)-\beta(i)}.\]
We use the $\ell_{\infty}$-metric to model a different kind of
noise-mechanism than that modeled by Kendall's $\tau$-metric,
namely spike noise. In this model, the rank of each memory cell is
assumed to have been changed by a bounded amount (see
\cite{TamSch10}).

Error-correcting and -detecting codes in $S_{n}$ for the
$\ell_{\infty}$-metric are referred to in \cite{TamSch10} as
\emph{limited-magnitude rank-modulation codes} (LMRM codes). In that
paper, constructions of such codes achieving non-vanishing normalized
distance and rate are presented. Moreover, bounds on the size of
optimal LMRM codes are proven. In particular, it has been shown
\cite[Th.~20]{TamSch10} that if $C$ is an $(n,M,2)$-LMRM then
\[M\leq\frac{n!}{2^{\floorenv{n/2}}}.\]
Using a simple translation to an extremal problem involving permanents
of $(0,1)$-matrices (see \cite{SchTam11}), this is also the best
possible bound using the set-antiset method. For our needs, it follows
that the size of every $n$-length $\ell_{\infty}$-snake is
bounded by this term. We shall present a construction of
$\ell_{\infty}$-snakes achieving this upper-bound by a factor of
$\floorenv{\frac{n}{2}}2^{\ceilenv{n/2}}$, which we will show achieves an asymptotic
rate of $1$.

\subsection{Construction}\label{infty_const}

In order to use the code constructions presented in
\cite{JiaMatSchBru09}, we first prove the following lemma.

\begin{lemma}\label{infty_codes_of_any_length}
Both constructions in \cite[Th.~4,7]{JiaMatSchBru09}, when applied
recursively, yield complete cyclic $n$-RMGC's containing both
\pttt operations $t_{2}$ and $t_{n}$.
\end{lemma}
\begin{IEEEproof}
The proposition was, while not fully stated, actually proven in \cite[Th.~4]{JiaMatSchBru09}.

For \cite[Th.~7]{JiaMatSchBru09}, we shall assume that the recursive
process was applied to a length-$(n-1)$ Gray code satisfying these
conditions (as is the case with the base example given in that
article). The resulting code uses $t_{n}$ by definition. Moreover,
since the original code used $t_{n-1}$, the resulting code uses
$t_{n-(n-1)+1}=t_{2}$.
\end{IEEEproof}

This lemma now allows for the construction of a basic building block
which we will later use.

\begin{lemma}\label{infty_aux_const}
Let $\mathset{a_{j}}_{j=1}^{n}$, $n\geq 2$, be a set of integers of the same
parity. Let
\[\sigma = \sparenv{x,a_{1},a_{2},\ldots,a_{n},b_{n+2},b_{n+3},\ldots,b_{m}}\in S_{m}\]
be a permutation such that the parity of $x$ differs from that of the
elements of $\mathset{ a_{j}}_{j=1}^{n}$. Then there exists a
(non-cyclic) $(m,n+(n-1)!,\ell_\infty)$-snake starting with $\sigma$
and ending with the permutation
\[t_{2} t_{n+1}^{n}(\sigma) = \sparenv{a_{2},a_{1},a_{3},a_{4},\ldots,a_{n},x,b_{n+2},b_{n+3},\ldots,b_{m}}.\]
\end{lemma}
\begin{IEEEproof}
Let $\sigma_0,\dots,\sigma_{n+(n-1)!-1}$ denote the codewords of the
claimed code, and denote by $t_{k_1},\dots,t_{k_{n+(n-1)!-1}}$ the
list of transformations generating it.

We set $\sigma_0=\sigma$. For all $i\in[n]$ we let $\sigma_{i}\triangleq
t_{n+1}^{i}(\sigma)$, i.e., $t_{k_i} = t_{n+1}$. Quite
clearly, any two of these $n+1$ permutations are at $\ell_{\infty}$-distance
at least $2$ apart, since the $a_{j}$'s share parity.

Now, by Lemma \ref{infty_codes_of_any_length} there exists a complete
cyclic $(n-1)$-RMGC starting with $\sigma_n$, with its last operation
being $t_{2}$. We therefore let $t_{k_{n+i}}$ for $i\in[(n-1)!]$
represent that code, hence $t_{k_{n+(n-1)!}} = t_{2}$ and
$\sigma_{n+(n-1)!} = \sigma_{n}$ (we then, obviously, omit the last
transformation as well as the repeated codeword
$\sigma_{n+(n-1)!}$). These $(n-1)!$ permutations,
$\sigma_n,\dots,\sigma_{n+(n-1)!-1}$, also represent an
$\ell_\infty$-snake, for the same reason.

Finally, take $0\leq k<n$ and $0\leq l<(n-1)!$, and observe
$\sigma_{k}$ and $\sigma_{n+l}$. Suppose
$d_{\infty}(\sigma_{k},\sigma_{n+l})\leq 1$. Then in particular
$\abs{a_{n-k}-x}=1$. Moreover, if $k=n-1$ then $\abs{x-a_{n}}=1$, but
then $a_{n}$'s position in $\sigma_{k}$ correlates to one of
$\mathset{ a_{j}}_{j=1}^{n-1}$ in $\sigma_{n+l}$, in
contradiction. Therefore $k\leq n-2$, but then $a_{n}$'s position in
$\sigma_{n+l}$ ($n$th from left) correlates to that of $a_{n-k-1}$ in
$\sigma_{k}$, where $1\leq n-k-1\leq n-1$, again in
contradiction. This concludes our proof.
\end{IEEEproof}

Having this building block in hand, we continue to describe a
construction of a cyclic $\ell_\infty$-snake. The construction follows
by dividing the ranks in a length-$n$ permutation into even and odd
elements, and covering permutations on each half separately.

\begin{theorem}\label{infty_code}
For all $4\leq n\in\N$ there exists an $(n,M,\ell_{\infty})$-snake
of size
\[M=\ceilenv{\frac{n}{2}}!\parenv{\floorenv{\frac{n}{2}}+\parenv{\floorenv{\frac{n}{2}}-1}!}.\]
\end{theorem}
\begin{IEEEproof}
To simplify notations, we start by noting that $[n]$ has
$p\triangleq\ceilenv{\frac{n}{2}}$ odd elements and
$q\triangleq\floorenv{\frac{n}{2}}$ even ones. We shall use that
notation throughout this proof.

Using \cite[Th.~4,7]{JiaMatSchBru09} we take a complete cyclic
$p$-RMGC using the operations
\[t_{\alpha(1)},t_{\alpha(2)},\ldots,t_{\alpha(p!)}.\]
Moreover, we use Lemma \ref{infty_aux_const} to come by a
$(q,M_q,\ell_\infty)$-snake of size $M_q=q+\parenv{q-1}!$ given by the
operations
\[t_{\beta(1)},t_{\beta(2)},\ldots,t_{\beta(q+\parenv{q-1}!-1)}.\]

As the origin for the code we construct we use
\[\sigma_{0} \triangleq \sparenv{1,2,4,\ldots,2q,3,\ldots,2p-1}.\]
For all $i\in\sparenv{p!}$ and $j\in\sparenv{q+\parenv{q-1}!-1}$ we define
sequence of transformations generation the code as
\begin{align*}
t_{k_{(i-1)\parenv{q+\parenv{q-1}!}+j}} & \triangleq t_{\beta(j)}\\
t_{k_{i\parenv{q+\parenv{q-1}!}}} & \triangleq t_{\alpha(i)+q+1},
\end{align*}
and where, naturally, the codewords satisfy $\sigma_{i}=t_{k_{i}}(\sigma_{i-1})$.

We start by noting that, for all $i\in[p!]$, the permutation
$\sigma_{(i-1)\parenv{q+\parenv{q-1}!}}$ satisfies the requirements of
Lemma \ref{infty_aux_const} as a simple matter of induction. It
follows that for all $i\in\sparenv{p!}$ the permutations
\[\mathset{\sigma_{(i-1)\parenv{q+\parenv{q-1}!}+1},\sigma_{(i-1)\parenv{q+\parenv{q-1}!}+2},\ldots,\sigma_{i\parenv{q+\parenv{q-1}!}-1}}\]
are at $\ell_{\infty}$-distance of at least $2$ apart.

Furthermore, for $i,i'\in[p!]$, $i<i'$, since the code generated by
$t_{\alpha(1)},t_{\alpha(2)},\ldots,t_{\alpha(p!)}$
is indeed a Gray code, we are assured that for all $0\leq j,j'\leq
q+(q-1)!-1$ the last $p-1$ elements of both
$\sigma_{(i-1)\parenv{q+(q-1)!}+j}$ and $\sigma_{(i'-1)\parenv{q+(q-1)!}+j'}$
are all odd and represent two distinct permutations, hence
\[d_{\infty}\parenv{\sigma_{(i-1) \parenv{q+(q-1)!}+j},\sigma_{(i'-1)\parenv{q+(q-1)!}+j'} }\geq 2.\]

Finally, we note that
\[t_{\alpha(p!)}\parenv{\sigma_{p!\parenv{q+(q-1)!}-1}}=\sigma_{0},\]
since the code provided by
$t_{\alpha(1)},t_{\alpha(2)},\ldots,t_{\alpha(p!)}$
is cyclic and $o(t_{2})=2$ divides $p!$.
\end{IEEEproof}

We note that by switching the roles of odd and even numbers in Theorem
\ref{infty_code} we can construct an $(n,M,\ell_\infty)$-snake of size
\[M=\floorenv{\frac{n}{2}}!\parenv{\ceilenv{\frac{n}{2}}+\parenv{\ceilenv{\frac{n}{2}}-1}!}.\]
However, the resulting code is strictly smaller for odd $n$.

\begin{theorem}
The $\ell_\infty$-snakes constructed in Theorem \ref{infty_code} have
an asymptotically-optimal rate.
\end{theorem}

\begin{IEEEproof}
Let $C_n$ denote the $\ell_\infty$-snake of length $n$ constructed by
Theorem \ref{infty_code}. Using the crude
\[\parenv{\frac{n}{e}}^n \leq n! \leq n^n\]
the proof is a matter of simple calculation:
\begin{align*}
\lim_{n\rightarrow\infty}R(C_n) & = \lim_{n\rightarrow\infty} \frac{\log_2\parenv{\ceilenv{\frac{n}{2}}!\parenv{\floorenv{\frac{n}{2}}+\parenv{\floorenv{\frac{n}{2}}-1}!}}}{\log_2\parenv{n!}}\\
& \geq \lim_{n\rightarrow\infty} \frac{2\log_2\parenv{\parenv{\floorenv{\frac{n}{2}}-1}!}}{\log_2\parenv{n!}}\\
& \geq \lim_{n\rightarrow\infty} \frac{\parenv{n-4}\log_2\parenv{\frac{n-4}{2e}}}{n\log_2 n} = 1.
\end{align*}
\end{IEEEproof}

\subsection{Successor Calculation and Ranking Algorithms}

Finding the correct \pttt operation to propagate a given
permutation to the following one is naturally dependent upon one's ability
to do the same with the $\ceilenv{\frac{n}{2}}$- and 
$\parenv{\floorenv{\frac{n}{2}}-1}$-RMGC's used in our construction. We
therefore assume to have the function 
$\suc\parenv{\sparenv{a_{1},a_{2},\ldots,a_{n}}}$ which accepts as
input a permutation $\sparenv{a_{1},a_{2},\ldots,a_{n}}\in S_{n}$ and returns 
the correct transformation used in the codes we used.
Furthermore, we assume to have the function 
$\rn\parenv{\sparenv{a_{1},a_{2},\ldots,a_{n}}}$ which returns the 
respective rank of the input permutation in that code, where the identity 
permutation is assumed to have rank zero.
Finally, we shall use an auxiliary function $\sw:S_n\rightarrow S_n$ defined 
by $\sw\parenv{\sigma} \triangleq (1,2)\circ\sigma$ (which naturally operates in 
$O(n)$ steps).

The function $\lsucc\parenv{\sparenv{a_{1},\ldots,a_{n}}}$ then
returns as output the index $i$ of the required transformation $t_{i}$
to produce the subsequent permutation in the code from
$\sparenv{a_{1},\ldots,a_{n}}$. It operates by considering the
following cases: in each block of Lemma \ref{infty_aux_const} one
computes the proper index by propagating the leading element of odd
rank as long as that is needed, then applying $\suc$ to the
permutation on the elements of even ranks (where one distinguishes
between blocks in which $2,4$ were switched). Only the last
permutation of each block calls for applying $\suc$ to the permutation
on the elements of odd ranks.

\begin{function}[t]
\begin{footnotesize}
\DontPrintSemicolon
\Input{A permutation $[a_1,a_2,\ldots,a_n]$}
\Output{$i\in\mathset{2,3,\ldots,n}$ that determines the transition $t_i$ to the 
next permutation in the $\ell_\infty$-snake from Theorem \ref{infty_code}}
	$q\leftarrow\floorenv{\frac{n}{2}}$; $p\leftarrow\ceilenv{\frac{n}{2}}$\\
	\If {$a_{q+1}\equiv 0\pmod{2}$}
	{
		\Return $q+1$ \nllabel{alg:infty_successor:non-general_position}
	}
	\If {$\rn(\sparenv{\frac{a_{q+1}+1}{2},\ldots,\frac{a_{n}+1}{2}})\equiv 0 \pmod{2}$}
	{
		\If {$\sparenv{a_{1},\ldots,a_{q}} = \sparenv{4,2,6,\ldots,2q}$}
		{
			\Return $q+\suc\parenv{\sparenv{\frac{a_{q+1}+1}{2},\ldots,\frac{a_{n}+1}{2}}}$ \nllabel{alg:infty_successor:general_position:even:final}
		}
		\Return $\suc\parenv{\sparenv{\frac{a_{1}}{2},\ldots,\frac{a_{q}}{2}}}$ \nllabel{alg:infty_successor:general_position:even}
	}
	\If {$\sparenv{a_{1},\ldots,a_{q}} = \sparenv{2,4,\ldots,2q}$}
	{
		\Return $q+\suc\parenv{\sparenv{\frac{a_{q+1}+1}{2},\ldots,\frac{a_{n}+1}{2}}}$ \nllabel{alg:infty_successor:general_position:odd:final}\\
	}
	\Return $\suc\parenv{\sw\parenv{\sparenv{\frac{a_{1}}{2},\ldots,\frac{a_{q-1}}{2}}}}$ \nllabel{alg:infty_successor:general_position:odd}
\end{footnotesize}
\caption{() $\lsucc\parenv{\sparenv{a_1,\ldots,a_n}}$}
\label{alg:infty_successor}
\end{function}

\begin{lemma}\label{succ_complx}
If the functions $\suc,\rn$ operate in $L_n,M_n$ steps respectively in
the average case, then $\lsucc$ has an average run-time of
$O\parenv{n+L_{q-1}+M_p}$.
\end{lemma}
\begin{IEEEproof}
We partition our proof by return cases. $\lsucc$ exits at line
\ref{alg:infty_successor:non-general_position} in precisely
$\frac{q}{q+(q-1)!}$ of cases, in which case it returns within a fixed
number of operations.

It exits at lines
\ref{alg:infty_successor:general_position:even:final},
\ref{alg:infty_successor:general_position:odd:final} in
$\frac{1}{q+(q-1)!}$ of cases, in which case it operates in at most
(depending on the data structures in use) $O(n)+M_p+L_p$ steps in the
average case.

Finally, $\lsucc$ returns from lines
\ref{alg:infty_successor:general_position:even},
\ref{alg:infty_successor:general_position:odd} in
$\frac{(q-1)!-1}{q+(q-1)!}$ of cases, after performing
$O(n)+M_p+L_{q-1}$ steps.

In every sensible implementation of $\suc$ (i.e., where we assume
$\frac{L_p-L_{q-1}}{q+(q-1)!}\rightarrow 0$) we then have an amortized run-time of
$O\parenv{n+L_{q-1}+M_p}$.
\end{IEEEproof}

We now note that by \cite[Th.~7,10]{JiaMatSchBru09} we may assume
$\suc$ to operate in $O(1)$ steps in the average case, and by
\cite[Part III-C]{JiaMatSchBru09} (which also relies on
\cite{MarStr07}) we assume $\rn$ runs in $O(n)$ steps, yielding an
average run-time of $O(n)$ for $\lsucc$.

We shall also present the function $\lrank(n,[a_{1},\ldots,a_{n}])$
that, given a permutation in the $\ell_\infty$-snake presented in part
\ref{infty_const}, returns that permutation's rank in the code. This
function uses the function $\rn$ discussed above as well, and works by
considering the same cases discussed above.

\begin{function}[t]
\begin{footnotesize}
\DontPrintSemicolon
\Input{A permutation $\sparenv{a_1,a_2,\ldots,a_n}$ in the $\ell_\infty$-snake 
from Theorem \ref{infty_code}}
\Output{$k\in\N$ that represents the given permutation's rank in the code}
	$q\leftarrow\floorenv{\frac{n}{2}}$; $p\leftarrow\ceilenv{\frac{n}{2}}$\\
	\If {$a_{q+1}\equiv 0\pmod{2}$}
	{
		$i\leftarrow\min\mathset{j\in[n] ~|~ a_j \not\equiv 0 \pmod{2}}$\\
		\Return $i-1+(q+(q-1)!)\cdot\rn\parenv{\sparenv{\frac{a_{i}+1}{2},\frac{a_{q+2}+1}{2},\ldots,\frac{a_{n}+1}{2}}}$ \nllabel{alg:infty_rank:non-general_position}
	}
	$R\leftarrow \rn\parenv{\sparenv{\frac{a_{q+1}+1}{2},\ldots,\frac{a_{n}+1}{2}}}$\\
	\If {$R\equiv 0\pmod{2}$}
	{
		\Return $q+(q+(q-1)!)\cdot R + \rn\parenv{\sparenv{\frac{a_{1}}{2},\ldots,\frac{a_{q-1}}{2}}}$ \nllabel{alg:infty_rank:general_position:even}
	}
	\Return $q+(q+(q-1)!)\cdot R + \rn\parenv{\sw\parenv{\sparenv{\frac{a_{1}}{2},\ldots,\frac{a_{q-1}}{2}}}}$ \nllabel{alg:infty_rank:general_position:odd}\\
\end{footnotesize}
\caption{() $\lrank\parenv{\sparenv{a_1,\ldots,a_n}}$}
\label{alg:infty_rank}
\end{function}

\begin{function}[t]
\begin{footnotesize}
\DontPrintSemicolon
\Input{$4\leq n\in\N$; rank $k\in\N$}
\Output{The permutation $[a_{1},a_{2},\ldots,a_{n}]$ which is $k$th in the 
$(n,M,\ell_\infty)$-snake from Theorem \ref{infty_code}}

	$q\leftarrow\floorenv{\frac{n}{2}}$; $p\leftarrow\ceilenv{\frac{n}{2}}$\\
	$R\leftarrow\floorenv{\frac{k}{q+(q-1)!}}$; $r\leftarrow\parenv{k\bmod(q+(q-1)!)}$\\
	$\sparenv{b_{1},\ldots,b_{p}}\leftarrow\texttt{UnR}(p,R)$\\
	\If {$r\geq q$}
	{
		$\sparenv{a_{1},\ldots,a_{q-1}}\leftarrow\unr(q-1,r-q)$\\
		\If {$R\equiv 1 \pmod{2}$}
		{
			$\sparenv{a_{1},\ldots,a_{q-1}}\leftarrow\sw\parenv{\sparenv{a_{1},\ldots,a_{q-1}}}$
		}
		\Return $[2a_{1},\ldots,2a_{q-1},2q,2b_{1}-1,\ldots,2b_{p}-1]$ \nllabel{alg:infty_unrank:general_position}
	}
	\If {$R\equiv 0 \pmod{2}$}
	{
		\Return \begin{tiny} $\sparenv{2,4,6,\ldots,2r,2b_{1}-1,2(r+1),\ldots,2q,2b_{2}-1,\ldots,2b_{p}-1}$\end{tiny} \nllabel{alg:infty_unrank:non-general_position:even}
	}
		\Return \begin{tiny} $\sparenv{4,2,6,\ldots,2r,2b_{1}-1,2(r+1),\ldots,2q,2b_{2}-1,\ldots,2b_{p}-1}$\end{tiny} \nllabel{alg:infty_unrank:non-general_position:odd}

\end{footnotesize}
\caption{() $\lunrank\parenv{n, k}$}
\label{alg:infty_unrank}
\end{function}

\begin{lemma}\label{rank_complx}
If the function $\rn$ operates in $M_n$ steps, then $\lrank$ has a run-time of 
$O(n+M_p)$ (in the average or worst case respectively).
\end{lemma}
\begin{IEEEproof}
We partition our proof by return condition once more. If the program exits 
from \ref{alg:infty_rank:non-general_position} then it performed $O(q)+M_p$ steps.

If it exits from \ref{alg:infty_rank:general_position:even} or 
\ref{alg:infty_rank:general_position:odd} then it performed $O(1)+M_p+M_{q-1}$ steps.
\end{IEEEproof}

Again, by results discussed above, we note that $\lrank$ runs in
$O(n)$ steps in the average case.

\begin{figure*}[t]
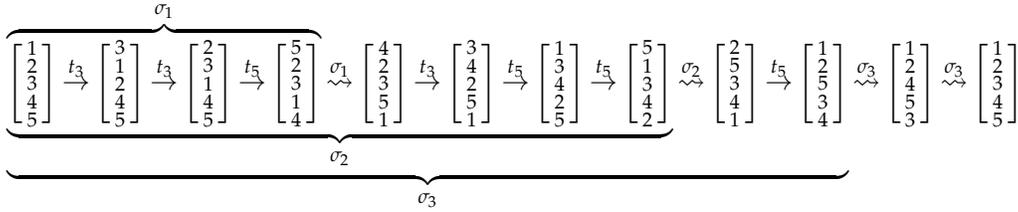

\begin{align*}
\underbrace{\underbrace{\overbrace{
\sparenv{\begin{smallmatrix} 1\\ 2\\ 3\\ 4\\ 5\end{smallmatrix}}
\operatorname*{\rightarrow}^{t_3}  \sparenv{\begin{smallmatrix} 3\\ 1\\ 2\\ 4\\ 5\end{smallmatrix}}
\operatorname*{\rightarrow}^{t_3}\sparenv{\begin{smallmatrix} 2\\ 3\\ 1\\ 4\\ 5\end{smallmatrix}}
\operatorname*{\rightarrow}^{t_5}\sparenv{\begin{smallmatrix} 5\\ 2\\ 3\\ 1\\ 4\end{smallmatrix}}}^{\sigma_1}
\operatorname*{\rightsquigarrow}^{\sigma_1}\sparenv{\begin{smallmatrix} 4\\ 2\\ 3\\ 5\\ 1\end{smallmatrix}}
\operatorname*{\rightarrow}^{t_3}\sparenv{\begin{smallmatrix} 3\\ 4\\ 2\\ 5\\ 1\end{smallmatrix}}
\operatorname*{\rightarrow}^{t_5}\sparenv{\begin{smallmatrix} 1\\ 3\\ 4\\ 2\\ 5\end{smallmatrix}}
\operatorname*{\rightarrow}^{t_5}\sparenv{\begin{smallmatrix} 5\\ 1\\ 3\\ 4\\ 2\end{smallmatrix}}}_{\sigma_2}
\operatorname*{\rightsquigarrow}^{\sigma_2}\sparenv{\begin{smallmatrix} 2\\ 5\\ 3\\ 4\\ 1\end{smallmatrix}}
\operatorname*{\rightarrow}^{t_5}\sparenv{\begin{smallmatrix} 1\\ 2\\ 5\\ 3\\ 4\end{smallmatrix}}}_{\sigma_3}
\operatorname*{\rightsquigarrow}^{\sigma_3}\sparenv{\begin{smallmatrix} 1\\ 2\\ 4\\ 5\\ 3\end{smallmatrix}}
\operatorname*{\rightsquigarrow}^{\sigma_3}\sparenv{\begin{smallmatrix} 1\\ 2\\ 3\\ 4\\ 5\end{smallmatrix}}
\end{align*}
\caption{ A $(5,57,\cK)$-snake generated by a computer
  search. Squiggly arrows stand for a repetition of the transitions
  defined by the braces. }
\label{ex:fullksnake}
\end{figure*}

It may prove important to identify the permutation associated with a
specific rank in our code. For that purpose we implement the function
$\lunrank(n,k)$, accepting as input the length of the code and a
specific rank and returning the implied permutation. We will assume
the existence of a similar function $\unr$ for the construction used
in part \ref{infty_const}, where again we assume the unit permutation
to have rank zero.

Once more, our implementation and estimate of $\lunrank$'s run-time
relies heavily on that of its auxiliary functions.

\begin{lemma}
If the function $\unr$ operates in $N_n$ steps, then $\lunrank$ runs in $O(n+N_p)$ steps.
\end{lemma}
\begin{IEEEproof}
One notes that the only operations in $\lunrank$ that take more than a
fixed number of steps are calls for $\sw$ (taking $O(n)$), calls for $\unr$,
and, depending on the data structures in use, concatenation of
indices (at most $O(n)$ as well). The claim follows.
\end{IEEEproof}

Again, it shall be noted that, relaying on Lemma
\ref{infty_codes_of_any_length} and \cite[Part III-C]{JiaMatSchBru09},
$\lunrank$ can be performed in $O(n^{2})$ operations.

\section{Conclusion}\label{conclusion}

In this paper we explored rank-modulation snake-in-the-box codes under
both Kendall's $\tau$-metric and the $\ell_\infty$-metric. In both
cases we presented a construction yielding codes with
asymptotically-optimal rates, and implemented auxiliary functions for
the production of the successor permutation, as well as ranking and
unranking for permutations in such codes. We also proved upper-bounds
on the size of $\cK$-snakes.

However, it is not presently known whether the upper-bounds presented
and referenced in this paper are achievable.  A computer search for
\emph{cyclic} codes, performed on $S_5$, yielded $(5,M,\cK)$-snakes of
maximal size $M=57$ (for comparison, the construction from Theorem
\ref{ksnakes} yields a $(5,45,\cK)$-snake).  While an abundance of
such codes were found (well over $500$ nonequivalent codes), they all
were in fact codes over $A_5$. For completeness, we present one of
those codes in Fig.  \ref{ex:fullksnake}.

Searches of a higher order appear to be infeasible, but we include one
more peculiar result: every maximal code we tested skipped $3$
permutations who all agree on $4,5$, i.e., it skipped a coset of
$S_3$. While we have no optimal codes of a higher order to test this
phenomenon on, the codes generated by Theorem \ref{ksnakes} of lengths
$7$ and $9$ display it as well - several cosets of $S_5$ and $S_7$
were absent, respectively.

It shall be noted that a complete (but not cyclic) $(5,60,\cK)$-snake
over $A_5$ can easily be constructed from each cyclic code we tested by
generating the skipped coset of $S_3$ with two $t_3$ operations,
followed by a $t_5$ operation and the given code, in order. However, 
we do not currently know whether $(2n+1,\frac{(2n+1)!}{2},\cK)$-snakes 
over $A_{2n+1}$ exist for every length.

These results, along with the bounds we showed in Lemmas
\ref{K-t_odd_paving} and \ref{K-t_full_paving} give rise to the
following conjecture: For all $n\in\N$ a $\cK$-snake exists over $A_n$
whose size is no less than that of every $\cK$-snake over $S_n$.

In addition, searches done in a computer for $\ell_\infty$-snakes for
lengths $4,5,6$ returned codes of size $6,30,90$ respectively,
suggesting that perhaps the upper-bound of \cite[Th.~20]{TamSch10} is
achievable. Moreover, in these cases we were able to find codes
generated only by \pttt operations on the last two indices. A code for
each length is presented in Fig. \ref{ex:ellsnakes} in binary
representation (conveniently written in octal notation), where zeroes
stand for $t_n$'s and ones for $t_{n-1}$'s. Searches for higher
lengths again seem infeasible.

\begin{figure}[t]
\[\begin{array}{c|l}
n & \text{Defining Transitions} \\
\hline\hline
4 & \mathtt{55} \\
5 & \mathtt{0212206063} \\
6 & \mathtt{010204410222042124446130162347}
\end{array}\]
\caption{ $(4,6,\ell_\infty)$-, $(5,30,\ell_\infty)$- and
  $(6,90,\ell_\infty)$-snakes generated by a computer search. All
  codes represented by a sequence of \pttt operations, applied in
  order to the identity permutation, where zeroes stand for $t_n$'s
  and ones for $t_{n-1}$'s. The binary strings are given in
  octal notation and should be read from left to right.}
\label{ex:ellsnakes}
\end{figure}

\bibliographystyle{IEEEtranS}
\bibliography{allbib}

\end{document}